\documentclass[doublecol]{epl2} 

\usepackage{bm,graphicx,amsmath}
\usepackage{bbm}

\usepackage{color}
\usepackage{amssymb}
\usepackage{amsmath}
\usepackage{amstext}
\usepackage{latexsym}
\usepackage{dsfont}

\newcommand{\ave}[1]{\left\langle#1\right\rangle}

\newcommand{\tr}[1]{{\rm Tr}\{#1\}}

\newcommand{\XX}{X} 
\newcommand{\PP}{P} 
\newcommand{\xx}{x} 
\newcommand{\pp}{p} 

\title{Entangling two distant oscillators with a quantum reservoir}

\author{Alexander Wolf\inst{1} \and Gabriele De Chiara\inst{2,3} \and Endre Kajari\inst{1,4} \and Eric Lutz\inst{5} \and Giovanna Morigi\inst{2,4}}
\shortauthor{A. Wolf, G. De Chiara, E. Kajari, E. Lutz and G. Morigi}

\institute{                    
  \inst{1} Institute of Quantum Physics, Ulm University, D-89069 Ulm, Germany\\
  \inst{2} Grup d'{\`O}ptica, Departament de F{\'i}sica, Universitat Aut{\`o}noma de Barcelona, E-08193 Bellaterra, Spain\\
  \inst{3} F{\'i}sica Te\`orica: Informaci\'o i Processos Qu\`antics, Universitat Aut{\`o}noma de Barcelona, E-08193 Bellaterra, Spain\\
  \inst{4} Theoretische Physik, Universit{\"a}t des Saarlandes, D-66041 Saarbr{\"u}cken, Germany\\
  \inst{5} Department of Physics, University of Augsburg, D-86135 Augsburg, Germany\\
}
\pacs{03.67.Bg}{Quantum information, entanglement production}
\pacs{03.65.Yz}{Decoherence, quantum mechanics}
\pacs{05.40.Ca}{Noise, fluctuation phenomena}

\abstract{The generation of entanglement between two oscillators that interact via a common reservoir is theoretically studied. The reservoir is modeled by a one-dimensional harmonic crystal initially in thermal equilibrium. Starting from a separable state, the oscillators can become entangled after a transient time, that is of the order of the thermalization time scale. This behaviour is observed at finite temperature even when the oscillators are at a distance significantly larger than the crystal's interparticle spacing. The underlying physical mechanisms can be explained by the dynamical properties of the collective variables of the two oscillators which may decouple from or be squeezed by the reservoir. Our predictions can be tested with an ion chain in a linear Paul trap.}

\begin{document}

\maketitle

\section{Introduction}
Dissipation and decoherence seriously limit the possibility to observe quantum effects in the macroscopic world. Understanding their origin and dynamics at the microscopic level is a fundamental issue for quantum technologies, which aim at achieving control of the dynamics of scalable quantum systems. Microscopic models based on Hamiltonian dynamics have been studied over the last century. A textbook example is Brownian motion~\cite{RMP_Chandrasekhar,weiss1999}: damping and thermalization of a particle's motion is here found when the particle is embedded in a crystalline structure, which acts as a thermal bath under specific conditions~\cite{ford1965,rubin1963,weiss1999}. The physical picture obtained from these models supports the understanding that loss of coherence emerges from the creation of quantum correlations between system and reservoir~\cite{Eisert,hilt2009,Zurek}.\\
\indent
Recent works pointed out that the interaction with a common external reservoir may endorse the creation of entanglement between two physical systems, for instance two spins~\cite{braun2002,mundarain2007,venuti2006,bellomo2008,guinea2009} or two oscillators~\cite{guinea2009,audenaert2002,huelga2002,plenio2004,prauzner-bechcicki2004,anders2008,paz2008,zell2009} which are not directly coupled to each other. These studies show that the bath can exert an active role in establishing entanglement. In most cases, this observation can be explained using symmetry reasons: a collective variable of the systems is {\it decoupled} from the reservoir such that the corresponding eigenstates form a Decoherence-Free Subspace (DFS)~\cite{Lidar}. Provided that the systems initial states and the reservoir temperature fulfil certain conditions, entanglement can then be found at time scales at which a single system would have otherwise thermalized. A different mechanism for entanglement generation is based on the direct {\it coupling} of a collective variable with the bath~\cite{paz2008} and results from the squeezing of its variance at sufficiently low temperatures of the reservoir~\cite{weiss1999,hilt2009}.\\
\indent The dependence of entanglement when the systems are coupled to distant positions of the reservoir has recently been object of several studies~\cite{venuti2006,bellomo2008,guinea2009,audenaert2002,anders2008,zell2009}. In the thermodynamic limit, entanglement is found between a pair of oscillators of a one-dimensional chain only when these are nearest-neighbours~\cite{audenaert2002,anders2008}. In ref.~\cite{zell2009}, it was argued that entanglement between two oscillators which couple to a bath described by a generalised Caldeira-Leggett model~\cite{caldeira1983} disappears at distances larger than the wavelength associated with the cutoff frequency of an Ohmic spectral density~\cite{weiss1999}, which corresponds to the interparticle spacing when the reservoir is a crystalline structure. From these works one may expect that this length constitutes a limit, above which entanglement cannot be established by a reservoir of arbitrarily large size.

\begin{figure}[h]
\centering
\includegraphics[width=0.75\columnwidth]{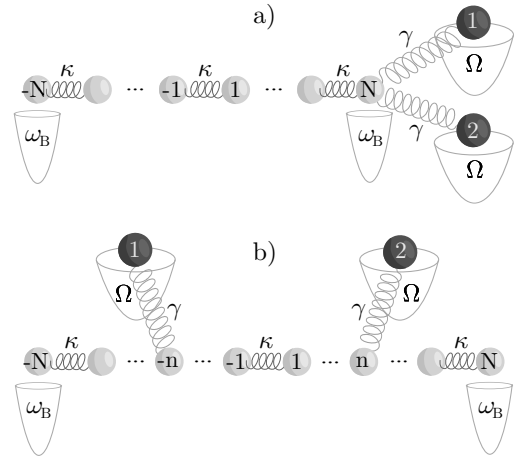}
\caption{Microscopic models a) and b) used to study entanglement creation between two oscillators 1 and 2 (dark grey) that are confined to a harmonic trap with frequency~$\Omega$. The oscillators couple linearly (with strength $\gamma$) to a harmonic chain that consists of $2N$ particles (light grey). The ions of the chain interact via nearest-neighbour coupling at strength $\kappa$. The traps with frequency $\omega_B$ pin the edge oscillators labeled by~$-N$ and~$N$.}
\label{fig:MicroModel}
\end{figure}

In this Letter we show that these expectations are to a large extent unfounded. We analyse the creation of entanglement between two harmonic oscillators that interact via a common reservoir --- a chain of harmonic oscillators with nearest-neighbour coupling. We first examine the conditions for which the chain acts as an Ohmic bath, in the setup of fig.~\ref{fig:MicroModel}a) which is analogous to Rubin's model when a single oscillator couples to the chain~\cite{weiss1999,rubin1963}. In this particular case, our results are in qualitative agreement with ref.~\cite{paz2008}, where the bath was described in terms of an Ohmic spectral density. We then demonstrate that the same reservoir can also support the creation of entanglement at long times between oscillators which couple to different, distant particles of the chain, as depicted in fig.~\ref{fig:MicroModel}b).

\section{Hamiltonian dynamics and initial states}
Our model consists of two oscillators with mass~$M$, frequency $\Omega$, position $\XX_\sigma$ and momentum $\PP_\sigma$ ($\sigma=1,2$) that interact with a particle of a linear chain. The linear chain is composed by $2 N$~particles of mass $m$, position $x_j$ and momentum $p_j$ ($j=\pm 1,\ldots, \pm N$) that couple via nearest-neighbour interaction. The edge particles are pinned by a harmonic trap with frequency~$\omega_B$. The total Hamiltonian ${H=H_S+H_B+H_I}$ comprises the Hamiltonians of the two oscillators, the chain, and their mutual interaction:
\begin{eqnarray}
&&H_S=\sum_{\sigma=1}^2 \left[ \frac{\PP^2_\sigma}{2 M} + \frac{1}{2}M \Omega^2 \XX^2_\sigma \right]\,,
\label{eq:HS}\\
&&\nonumber 
H_B=\sum_{i=-N}^{N}\left[\frac{\pp_i^2}{2 m}+\frac{m}{2}\omega_i^2\, \xx_i^2\right]
+\frac{\kappa}{2}\sum_{i=-N}^{N-1}(\xx_i- \xx_{i+1})^2\,,\\
&&H_I=\frac\gamma 2 \Big[(\XX_1-\xx_j)^2+(\XX_2-\xx_{k})^2\Big]\Theta(t)\,.
\label{eq:HI}
\end{eqnarray}
Here, $\kappa$ and $\gamma$ describe the coupling strengths and ${\omega_i=\omega_B\left(\delta_{i,-N}+\delta_{i,N}\right)}$ sets appropriate boundary conditions on the chain. The Heaviside function $\Theta(t)$ indicates that $H_I$ vanishes before the time $t=0$. For later convenience, we introduce the cutoff frequency~$\omega_{\rm cut}=\sqrt{4\kappa/m}$ of the chain's normal modes and denote by $a$ the interparticle distance, with $2a$ being the wavelength of the mode at frequency $\omega_{\rm cut}$~\cite{rubin1963}. Moreover, we define a reference frequency $\Omega_0$ such that $\Omega=(1+\epsilon)\,\Omega_0$ with a dimensionless parameter~$\epsilon$ that will be specified later.

The dynamics of the composite system is evaluated numerically, assuming that each oscillator is initially prepared in a squeezed vacuum state of Hamiltonian~(\ref{eq:HS}). We denote by $r$ the squeezing parame\-ter which is taken to be real such that the variances are $\Delta \XX^2_\sigma(0)={\rm e}^{-2r}\alpha^2/2$ and ${\Delta \PP^2_\sigma(0)={\rm e}^{2r}\hbar^2/(2\alpha^2)}$ with ${\alpha=\sqrt{\hbar/(M\Omega_0)}}$. The chain is prepared in a thermal state at temperature~$T$, and its density matrix reads $\rho_B(0)={\rm e}^{-\beta H_B}/Z$ with $Z=\tr{{\rm e}^{-\beta H_B}}$ and $\beta=(k_B T)^{-1}$. By integrating the equations of motion of the composite system, we determine the time-dependent covariance matrix ${V'_{\mu\nu} = \frac 12\langle \xi'_\mu \xi'_\nu + \xi'_\nu \xi'_\mu\rangle -\langle \xi'_\mu\rangle\langle \xi'_\nu\rangle}$ (${\mu, \nu\in\{1,2,3,4\}}$) in terms of the dimensionless variables of the oscillators ${\xi'_\mu=(X'_1,P'_1,X'_2,P'_2})$ with ${X'_\sigma=X_\sigma/\alpha}$ and ${P'_\sigma=P_\sigma\,\alpha/\hbar}$. The entanglement between the oscillators is evaluated by means of the logarithmic negativity 
${E_N = \max\{0,-\ln (2\tilde\nu_-)\}}$, where $\tilde\nu_{-}$ is the smallest symplectic eigenvalue of the partially transposed covariance matrix ${\tilde V = \Lambda V' \Lambda}$ that results from a mirror reflection ${\Lambda={\rm diag}(1,1,1,-1)}$ with respect to $P'_2$~\cite{vidal2002}.

\section{Microscopic model of an Ohmic bath}
We first focus on the model depicted in fig.~\ref{fig:MicroModel}a) where both oscillators couple to one edge of the chain, such that $j=k=N$. A simple transformation shows that only the centre-of-mass (COM) motion of the two oscillators with coordinate ${\XX_{+}=(\XX_{1}+\XX_{2})/\sqrt{2}}$ couples with particle $N$, while the relative motion with coordinate ${\XX_{-}=(\XX_{1}-\XX_{2})/\sqrt{2}}$ is decoupled. The coupling of the COM to the chain is analogous to Rubin's model of a single defect in a crystal~\cite{rubin1963}. Thermalization of the defect is expected when the crystal acts as an Ohmic bath. In order to recover this behaviour, we study the spectral density of the chain~\cite{caldeira1983,weiss1999}
\begin{equation}
J_{+}(\omega) = \frac{\pi}{2 m} \sum_{i=-N}^N \frac{\overline{\gamma}_i^2}{\overline{\omega}_i}\,\delta(\omega-\overline{\omega}_i)\,,
\end{equation}
where $\overline{\gamma}_i$ is the strength of the coupling between the COM oscillator and the chain's normal mode at frequency $\overline{\omega}_i$, whereby the normal modes diagonalize the quadratic Hamiltonian $H_B+ \gamma x_N^2$.

\begin{figure}[h]
\includegraphics[width=\columnwidth]{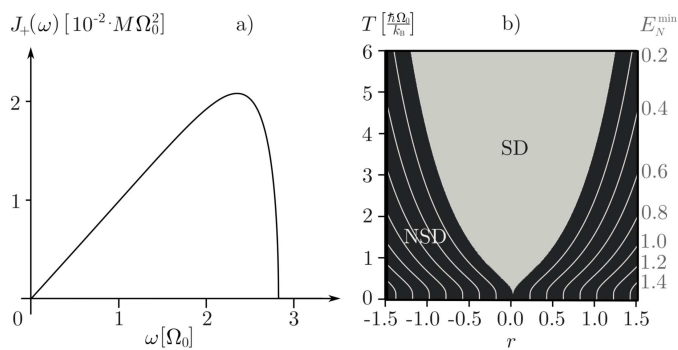}
\caption{a) Spectral density $J_+(\omega)$ for the model in fig.~\ref{fig:MicroModel}a), and b)
contour plot of the logarithmic negativity $E_N(r,T)$ at long times (but smaller than the revival time) as a function of the squeezing parameter~$r$ and temperature~$T$ (in units of $\hbar\Omega_0/k_B$). The black region (NSD) indicates the parameter regime in which the oscillators are entangled
(the SDR phase is not visible in this graph, but located between SD and NSD). The para{\-}meters are $2N=2500$, ${\epsilon=0}$, ${m=M/2}$, ${\omega_B=\sqrt{2}\,\Omega_0}$, ${\kappa=M\Omega_0^2}$, ${\gamma=0.1M\Omega_0^2}$ and $\omega_{\rm cut}=2\sqrt{2}\,\Omega_0$.}
\label{fig:rTPlot:dzero}
\end{figure}

By choosing the masses and the parameters $\kappa$, $\gamma$ and $\omega_B$ according to the prescriptions discussed in~\cite{rubin1963}, we find ${J_+(\omega)\sim \omega}$ for $\omega\lesssim \Omega$, which corresponds to the spectral density of an Ohmic bath and is displayed in fig.~\ref{fig:rTPlot:dzero}a). The thermalization process is numerically investigated by analysing the behaviour of the covariance matrix for times smaller than the revival time ${t_{\rm rev}\approx L/c_s}$, where ${L=2Na}$ and $c_s\approx\omega_{\rm cut}\,a/2$ are the length and the sound velocity of the chain, respectively. Note that $t_{\rm rev}$ grows linearly with~$N$. For ${2N=2500}$ and $\gamma=0.1\, M\Omega_0^2$, we observe that all correlations between COM and relative coordinates vanish at sufficiently long times $t$ (but $t<t_{\rm rev}$), and the only non-vanishing components of the covariance matrix are the {\it constant} widths $\Delta X_+^2$, $\Delta P_+^2$, and the elements $\Delta X_-^2$,
$\Delta P_-^2$ and ${\frac{1}{2}\ave{X_-P_-+P_-X_-}}$. The latter oscillate at the frequency~$2\Omega_{\gamma}$ where ${\Omega_{\gamma}=\eta\,\Omega}$ is the frequency of the oscillator for the relative motion which entails the frequency shift due to the interaction Hamiltonian~$H_I$ through the parameter~$\eta=\sqrt{1+\gamma/(M\Omega^2)}$. In accordance, the quantity $\mathcal{E}_N(t)=-\ln(2\tilde\nu_-(t))$ oscillates with~$2\Omega_{\gamma}$ between the values $\mathcal{E}^{\rm min}_N<\mathcal{E}^{\rm max}_N$.

Depending on the initial squeezing parameter $r$ and the temperature~$T$, we encounter three different situations for the entanglement of the two oscillators when the COM motion has reached its ``stationary'' state. Our findings qualitatively agree with the results of ref.~\cite{paz2008}, where this behaviour was predicted by means of a non-Markovian master equation with an Ohmic (sub-Ohmic, super-Ohmic) spectral density~\cite{Hu}. The different entanglement behaviours are summarised in fig.~\ref{fig:rTPlot:dzero}b) where the logarithmic negativity is displayed as a function of~$r$ and~$T$. They are indicated using the nomenclature introduced in~\cite{paz2008}: (i) entanglement {\it sudden-death} (SD)~\cite{Eberly} when any transient entanglement disappears, such that $\mathcal{E}^{\rm max}_N<0$, (ii) entanglement {\it sudden death and revival}~(SDR) when $\mathcal{E}^{\rm min}_N<0<\mathcal{E}^{\rm max}_N$, and (iii) entanglement {\it no sudden death}~(NSD) when $\mathcal{E}^{\rm min}_N>0$.

Entanglement between the two oscillators is found when the quadratures of the COM and relative motion exhibit two-mode squeezing~\cite{EPR}. An analysis of the logarithmic negativity at long times shows that the oscillators are entangled when either inequality (I)~${\Delta X'^2_+(T)<\frac{1}{2\eta}{\rm e}^{2|r-r_S|}}$ or (II)~${\Delta P'^2_+(T)< \frac{\eta}{2}{\rm e}^{-2|r-r_S|}}$ is satisfied, providing a relation between~$T$ and~$r$. Here we also introduced the squeezing parameter\footnote{An initially squeezed state with squeezing parameter~$r_S$ corresponds to the ground state of an oscillator with trap frequency~$\Omega_\gamma$.} ${r_S=\frac{1}{2}\ln\eta}$. Two-mode squeezing has two different origins. The first one is connected to the DFS for the relative motion which leads to entanglement for large $|r|$ according to inequality (I). At sufficiently low temperatures and for $r\approx r_S$, however, $\Delta \XX_+$ is squeezed due to the coupling with the bath~\cite{weiss1999,hilt2009} and entanglement is found when (II) is fulfilled. The corresponding region is located where the two NSD-regimes meet in fig.~\ref{fig:rTPlot:dzero}b) (not visible for the chosen scale).

\section{Long-distance entanglement}
\begin{figure}[h]
\centering
\includegraphics[width=\columnwidth]{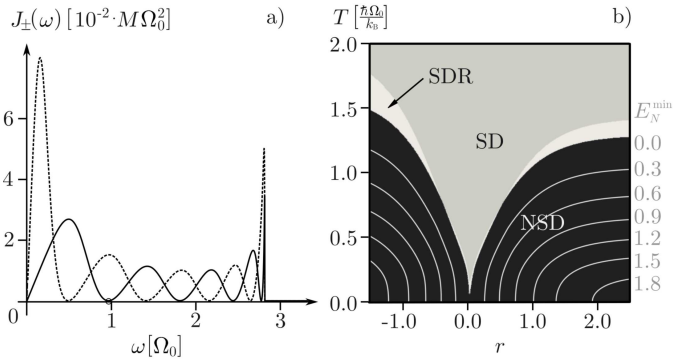}
\caption{a) Spectral densities $J_+(\omega)$ (dashed line) and $J_-(\omega)$ (solid line) corresponding to the centre-of-mass and relative motion of the oscillators, respectively, for the model of fig.~\ref{fig:MicroModel}b) with $d=9a$ and b) contour plot of the logarithmic negativity at long times as a function of $r$ and $T$. Here, $\epsilon=-0.086$, $2N=1500$; all other parameters are as in fig.~\ref{fig:rTPlot:dzero}.}
\label{fig:rTPlot:dnine}
\end{figure}

We now turn to the generation of entanglement when the oscillators are coupled to different particles of the reservoir, as illustrated in fig.~\ref{fig:MicroModel}b). Let the two particles of the chain be separated by a distance $d$ that is larger than the interparticle separation $a$. We set $j=-n$ and $k=n$ in eq.~(\ref{eq:HI}), such that $d=(2n+1)a$. After performing the transformation ${x^\pm_i=(x_i\pm x_{-i})/\sqrt2}$ (${i= 1,\ldots, N}$) for the bath particles, the whole dynamics reduces to the time evolution of two separate harmonic oscillators with coordinates~$\XX_\pm$ that interact with two independent reservoirs.
Figure~\ref{fig:rTPlot:dnine}a) displays the spectral density of the reservoir coupling with the centre-of-mass ($J_+(\omega)$) and the relative coordinate ($J_-(\omega)$). One observes that both spectral densities monotonically increase for frequencies $\omega$ sufficiently close to zero. For oscillator frequencies $\Omega$ that lie in this interval, the reservoir acts as an Ohmic environment for each collective variable. In this case, the oscillators reach a stationary state which exhibits no entanglement. This behaviour is found at any distance $d>a$ for sufficiently small values of $\Omega$ and $\gamma$, and is in agreement with the result of ref.~\cite{zell2009}, where the reservoir was described in terms of an Ohmic spectral density. 

A significantly different behaviour is obtained when the (shifted) oscillator frequency $\Omega_\gamma$ takes on values, for which the spectral density oscillates. In particular, when we choose $\epsilon$ such that $\Omega_\gamma$ coincides with a frequency value ${\omega=\omega^\pm_E}$ at which one of the spectral densities vanishes\footnote{The zeros of $J_\pm(\omega)$ read $\omega_{E,k}^{\pm}=\omega_{\rm cut}\,\sin\phi^{\pm}_k$ with ${\phi^{\pm}_k=\frac{\pi a}{2 d}(2k_{\pm}-\frac12\mp\frac12)}$, ${k_{\pm}\in\{1,\ldots,\lfloor \frac{d}{2a}\pm\frac14 \rfloor\}}$ and $\lfloor x \rfloor$ as floor function.}, ${J_\pm(\omega^\pm_E)=0}$, the corresponding variable $\XX_+$ or~$\XX_-$ decouples from the bath. In this case, entanglement at long times exhibits similar dynamics as for the model in fig.~\ref{fig:MicroModel}a), but with an important difference: the decoupled mode involves both the variable $\XX_\pm$ as well as the variables $x_i^\pm$ of the particles interposed between the oscillators.

Figure~\ref{fig:rTPlot:dnine}b) displays the logarithmic negativity at long times for ${d=9a}$, showing that long-distance entanglement occurs for sufficiently low temperatures~$T$.
Remarkably, we observe that the two oscillators cannot be entangled above a certain temperature $T$, independent of the initial squeezing $|r|$.  This is in sharp contrast to the action of the Ohmic environment (see fig.~\ref{fig:rTPlot:dzero}b)), where entanglement can be generated at any temperature, provided $|r|$ is sufficiently large. 

\begin{figure}[h]
\centering
\includegraphics[width=0.8\columnwidth]{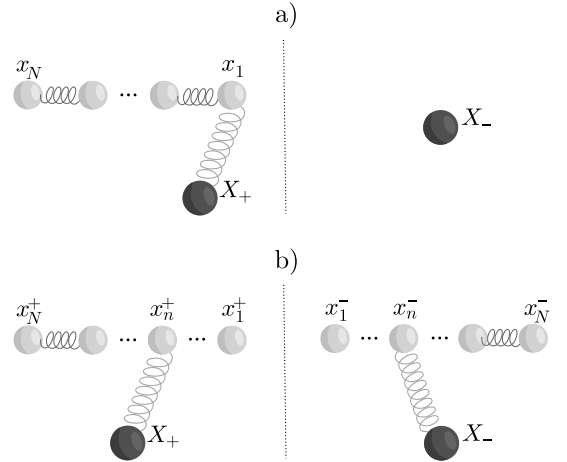}
\caption{a) Microscopic model of fig.~\ref{fig:MicroModel}a) and b) of fig.~\ref{fig:MicroModel}b) here reported in COM and relative coordinates ${\XX_{\pm}=(\XX_{1}\pm\XX_{2})/\sqrt{2}}$ of the two oscillators. In a), whereas the COM motion couples to the harmonic chain, the relative coordinate decouples from the rest of the dynamics. In b) COM and relative motion of the oscillators couple to two different reservoirs defined by the coordinates ${x_i^\pm=(x_{i}\pm x_{-i})/\sqrt{2}}$.}
\label{fig:COMandRel}
\end{figure}

The reason for this difference can be traced back to the type of DFS which corresponds to each model. In order to clarify it, let us recall the physical picture which explains entanglement in the model of fig.~1a). In this case, only the COM coordinate $X_+$ of the two oscillators interacts with the reservoir, whereas the relative coordinate $X_-$ is decoupled. Figure~\ref{fig:COMandRel}a) provides a sketch of the corresponding model in terms of collective variables. In this situation the squeezing of the initial state of the relative motion is preserved (even though the squeezed quadrature rotates at the oscillator frequency), while the correlations with the COM vanish. The relative motion will hence exhibit squeezing at long times, provided that the two oscillators are initially prepared in two independent squeezed states. This will lead to entanglement (two-mode squeezing) at any temperature as long as the initial squeezing parameter $r$ is sufficiently large. In the model of fig.~1b), in contrast, both the COM and relative coordinates of the two oscillators couple to a reservoir defined by the coordinates $x_i^+$ and $x_i^-$, respectively, with ${x_i^\pm=(x_{i}\pm x_{-i})/\sqrt{2}}$. However, in this case the two reservoirs are independent one from the other. This situation is illustrated in fig.~\ref{fig:COMandRel}b). Here, one can identify a mode, which involves $\XX_+$ ($\XX_-$) and the corresponding bath particles with coordinates $x_1^+,\ldots x_n^+$ ($x_1^-,\ldots x_n^-$), and that decouples from the rest of the chain provided that the spectral density $J_+(\omega)$ ($J_-(\omega)$) vanishes at the shifted oscillator frequency $\Omega_\gamma$. The initial state of this mode results from the initial squeezing of the corresponding collective variable of the two oscillators as well as from the thermal state of the involved chain oscillators. As a result, the width of the Gaussian state associated with the oscillator $\XX_+$ ($\XX_-$) of the decoupled mode depends also on the initial temperature of the chain. This property leads to the disapperance of long-distance entanglement above a certain temperature, irrespectively of the initial squeezing.

We further remark that our numerical simulations show that entanglement decreases linearly as the distance $d$ is increased. At fixed distances~$d$, however, the logarithmic negativity remains constant when the number~$N$ of particles in the chain is increased. Hence, entanglement at large distances can b found at arbitrarily long times in the thermodynamic limit.

Long-distance entanglement crucially relies on the non-Markovian nature of the bath and it appears when the trap frequency of the system particles is tuned to values at which the spectral density vanishes. Nonetheless, the behaviour of an Ohmic bath can be re-established by removing one of the system particles in fig.~\ref{fig:MicroModel}b), in which case the remaining oscillator thermalizes with the harmonic chain. This feature already highlights an important difference between our microscopic model and systems that are commonly employed for long-distance correlations such as dipoles that couple via a photonic-band gap material~\cite{Lambropoulos,bellomo2008}.

We note that the choice of our model is based on Rubin's harmonic crystal~\cite{rubin1963}, which serves as a paradigm for the analysis of quantum Brownian motion~\cite{weiss1999}. Unlike the systems studied in refs.~\cite{audenaert2002,anders2008} and \cite{zell2009}, where no distant entanglement is found, 
our model (i) possesses no discrete translational invariance and (ii) is characterised by a spectral density which is Ohmic only for a small frequency interval, whose width decreases as the oscillator's distance is increased. This shows in particular that approximating the environment with an Ohmic (sub-Ohmic, super-Ohmic) spectral density does not catch the relevant symmetries of the Hamiltonian which can support a DFS and thus entanglement creation. Deviations from this ideal behaviour, such as spatial inhomogeneity, are therefore expected to be detrimental as they destroy the symmetries of the coupling.

Our prediction could be tested with an ion string in a linear Paul trap, building on the proposals in~\cite{Retzker}. The oscillators could be identified with the radial vibrations of two heavier ions embedded in the chain, provided that they are sufficiently distant such that their mutual repulsion can be neglected. The axial and radial degrees of freedom of a defect correspond in our model to one oscillator and one particle of the chain, respectively. They can be harmonically coupled by means of dipolar potentials. With adequate radial trap frequencies and mass ratios between the defects and the other ions, the radial oscillators can be decoupled from the radial excitations of the rest of the chain. Moreover, their frequency could be tuned with respect to the axial spectrum by choosing appropriately the trap aspect ratio~\cite{Morigi04}.

\section{Summary}
We have studied a paradigmatic model based on Rubin's harmonic crystal which supports entanglement generation between two harmonic oscillators that indirectly couple via a one-dimensional harmonic chain. The creation of long-distance entanglement presented in this letter is neither a result of a controlled dynamical evolution, nor is it due to the finite size of the reservoir as in ref.~\cite{plenio2004}. Entanglement is found in the thermodynamic limit $N\to \infty$ and at finite temperatures (i) for distances between the oscillators that are significantly larger than the interparticle spacing of the chain and (ii) for times at which a single oscillator would have otherwise thermalized. These results may find useful applications in quantum technologies, for instance within the context of hybrid quantum networks based on quantum systems at the interface to a solid state environment.\\

\acknowledgments
It is a pleasure to thank T.~Calarco, C.~Cormick, M.~Palma, J.~P.~Paz, M.~Roncaglia, and especially M.~B.~Plenio and W.~P.~Schleich for fruitful discussions. Support by the European Commission (AQUTE; STREP PICC), by the Ministerio de Ciencia e Innovaci{\'o}n (QOIT Consolider-Ingenio 2010; FIS2007-66944; EUROQUAM; Juan de la Cierva), by the Generalitat de Catalunya (Grant No. 2005SGR-00343), by the German Research Foundation (LU1382/1-1, MO1845/1-1) and by the cluster of excellence ``Nanosystems Initiative Munich (NIM)'' is gratefully acknowledged.

\end{document}